\documentclass[journal,draftcls,onecolumn,11pt]{article}
 \usepackage{amssymb}
 \usepackage{graphicx}
  \usepackage{ifpdf}
  \pdfoutput=1
\begin{document}
\begin{center}
{\Large
Stability of routing strategies for the maximum lifetime problem in ad-hoc wireless networks}
\end{center}
\begin{center}
Z. Lipi{\'n}ski\\
Institute of Mathematics and Informatics, University of Opole
\end{center}

\begin{abstract}
We solve the maximum lifetime problem
for a one-dimensional, regular ad-hoc wireless network with one data collector $L_N$
for any data transmission cost energy matrix which elements $E_{i,j}$ are
superadditive functions, i.e., satisfy the inequality $\forall_{i\leq j\leq k}\;E_{i,j}+E_{j,k}\leq E_{i,k}$.
We analyze stability of the solution under modification of two sets of parameters,
the amount of data $Q_i$, $i\in [1,N]$ generated by each node
and location of the nodes $x_i$ in the network.
We assume, that the data transmission cost energy matrix $E_{i,j}$ is a function of a distance between network nodes
and thus the change of the node location causes change of $E_{i,j}$.
We say, that a solution $q(t_0)$ of the maximum network lifetime problem is stable under modification of
a given parameter $t_0$ in the stability region $U(t_0)$,
if the data flow matrix $q(t)$ is a solution of the problem for any $t\in U(t_0)$.
In the paper we estimate the size of the stability region $U(Q^0,d^0)$
for the solution of the maximum network lifetime problem for the $L_N$ network
in the neighborhoods of the points $Q^0_i=1$, $d^0_i=0$,
where $d_i\in (-1,1)$ describes the shift of the nodes from their initial location $x_i^0=i$, i.e., $x_i=i-d_i$.
\end{abstract}
Keywords: sensitivity analysis, wireless ad-hoc network, energy efficiency.
\section{Introduction}
Ad-hoc wireless networks are build of wireless nodes which activities
are related not only to prescribed functionality but also to the routing processes
and retransmission of data received from other nodes.
These activities are energy consuming and for networks which nodes characterize limited resources of energy
it is important to optimize its consumption.
This problem is especially important for networks which nodes cooperate to perform some computational tasks
and their activities should be designed in a way that the amount of energy consumed by each node
enables to perform these tasks for a long time.
There are several definitions of ad-hoc wireless network lifetime.
In \cite{Giridhar}, it was discussed the 'functional lifetime' of a sensor network,
defined as a number of cycles the network can perform its tasks before some node runs out of energy.
In \cite{Chang} the network lifetime $T_{{\rm Net}}$ was defined as the time until the first node drains out of energy,
and was given by the formula
$T_{{\rm Net}}(q) = \min_{i\in L_N} \{ \frac { E_{i}^{(0)}}{E_{i}^{}(q) } \}_{},$
where
$E_{i}^{(0)}$ is the initial energy of the $i$-th node and
$$E_{i}^{s}= \sum_{j=0, j\neq i}^{N} q_{i,j} E_{i,j}^{}$$
is the energy necessary
to send the amount $\sum_{j=1, j\neq i}^{N} q_{i,j}$ of data by the node to other nodes of the network.
The data flow matrix $q_{i,j}$ determines the amount of data which is sent from the $i$-th to the $j$-th node and the
matrix $E_{i,j}$ defines the cost of transmission of a one unit of data between these nodes.
To maximize the network lifetime it means to find such data flow matrix $q_{i,j}$,
that the objective function $T_{{\rm Net}}(q)$ of the problem reaches its maximum, i.e.,
$\max_{q} T_{{\rm Net}}(q).$
In \cite{Woo} it was identified five power-aware metrics for data transmission in ad-hoc networks
which can be used for definitions of the network lifetime.
The above two definitions are equivalent to the 'time to network partition',
the second metric discussed in \cite{Woo}.
In this paper we utilize the fifth metric defined in \cite{Woo},
and we will minimize the maximum node cost
$$\min_{q} \max_{i} \{ E_{i}^{s} \}.$$
For a one-dimensional, ad-hoc networks the metrics 'time to network partition'
and 'the maximum node cost'
are equivalent because the cut­set, a set of nodes the removal of which will cause the network to partition, \cite{Ford},
consists of all nodes of the network.
This means that obtained results in this paper are valid for
network lifetime problems discussed in \cite{Giridhar} and \cite{Chang}.

General formulation of the problem of extending the network lifetime depends on a number of initial parameters
which characterize the nodes activity, available resources, the network environment or topology.
It is known, that solutions of the problem are very sensitive under change of these parameters, \cite{Chang}.
In general even, if we know an exact solution of a given optimization problem
there is a need to investigate its stability under change of these initial parameters
to understand the structure of the solution and limitations of its applications, \cite{Griva}, \cite{Fiacco}.
In wireless ad-hoc networks which characterize limited power of energy,
for example in sensor networks, the nodes most of their energy utilize for data transmission and
extension of the network lifetime is achieved by designing energy efficient transmission protocols.
Knowledge about stability of the exact solution of the maximum network lifetime problem
allows to understand limitations of implemented transmission protocols in such networks.

In the paper we are considering stability of a solution of the maximum network lifetime problem for
a one-dimensional, regular ad-hoc wireless network with one data collector $L_{N}$.
The nodes of the regular network $L_N$ are
located at the points $x_i=i$, $i\in [0,N]$ of a half-line, where at the point $x_0=0$ there is
located a data collector.
We investigate stability of the solution under changes of two sets of parameters,
the amount of data $Q_i$, $i\in [1,N]$ generated by each node and location of the nodes $x_i$ in the network.
We assume, that the data transmission cost energy matrix $E_{i,j}$ is a function of a distance
between nodes of the network, i.e., $d_{i,j}=|x_i-x_j|$, and thus the change of the node location causes change of $E_{i,j}$.
The change of node location is described by the $d_i\in (-1,1)$ parameter which measures the shift of
the node from its initial location $x_i=i$.
In the paper we solve the maximum network lifetime problem for $L_N$ network for the data transmission cost energy matrix
having the superadditive property, \cite{Hille},
\begin{equation} \label{xi-xk-xj}
\forall_{x_i < x_j < x_k} \;\; E_{i,j} + E_{j,k} \leq E_{i,k}.
\end{equation}
Because the data transmission cost energy matrix $E_{i,j}=|x_i-x_j|^a$, where $a\geq 1$ satisfies (\ref{xi-xk-xj}),
then the most general form of $E_{i,j}$, which is an analytical function of a distance between nodes
of the $L_N$ network and has the property (\ref{xi-xk-xj}), can be given by the power series
\begin{equation} \label{Eij-a-n-lambda-n}
E_{i,j}(\bar{a},\bar{\lambda}) = \sum_{n=0}^{\infty} \lambda_{n} |x_i-x_j|^{a_n},
\end{equation}
where $\lambda_n\geq 0$, $a_n\geq 1$ and
$\forall_{i,j}\; 0\leq E_{i,j}(\bar{a},\bar{\lambda}) < \infty$.
For a regular ad-hoc network $L_N$ the distance between neighboring nodes is equal to one, $d_{i,i\pm 1}=1$
and we can assume, without loosing the generality of the problem that $E_{i,i\pm 1}(\bar{a},\bar{\lambda})=1$,
which is equivalent to the requirement $\sum_{n=0}^{\infty} \lambda_{n} =1$.
From the property (\ref{xi-xk-xj}) it follows, that the lowest costs of delivery $Q_{i}$ of data from the
$i$-th node located at $x_i$ to the $j$-th node located at $x_j$ is the transmission along the shortest path
with the 'next hop' transmission strategy.
By a shortest path data transmission between two nodes $x_i$ and $x_j$ we mean
a transmission along the distance $d(x_i,x_j)$.
By $S_N$ we denote a one-dimensional ad-hoc wireless network which nodes are located at arbitrary point $x_{i}$ of a half-line.
The network $S_N$ can be constructed from the regular one by change the location of the $L_N$ network nodes
to the points $x_i= i - d_i$, where $d_i\in (-1,1)$.
For a given solution $q(x^0)$ of the maximum lifetime problem for the $L_N$ network, where $x^0_i=i$,
we will search for a non-regular network $S_N$, with nodes located at the points $x_i= i - d_i$,
for which the solution $q(x^0)$ can be extended.
In other words, we will search for a region $U(d)$ for which the function $q(x^0-d)$,
$\forall d \in U(d)$ is a solution of the problem.
Generally, we say that a solution $q(d^0)$ of the maximum network lifetime problem
is stable under modification of the parameter $d^0$ in the region $U(d^0)\subset R^N$,
if for any $d \in U(d^0)$ the data flow matrix $q(d)$ is also a solution of the problem.
The region $U(d^0)$ we call the stability region of the solution $q(d)$ with respect to the parameter $d$.
In the next sections we analyze the scope of the parameters $Q_i$ and $d_i$, $i\in [1,N]$
for which the solution $q(Q^0,d^0)$ of the maximum network lifetime problem in $L_N$ network
can be continuously extended.
\section{Formulation of the problem}
The general definition of the maximum network lifetime problem for a one-dimensional,
ad-hoc wireless network with $N$ nodes and one data collector $S_{N}$ can be written
in the following form
\begin{equation} \label{minimaxNP}\left\{ \begin{array}{l}
\min_{q} \max_{i\in L_N} \{ E_{i}^{s} \}, \\
E_{i}^{s} = E_{i}^{(R)} \sum_{j \in V^{(in)}_{i}} q_{j,i} +  E_{i}^{(Q)} Q_{i} + \sum_{j \in V^{(out)}_{i}} q_{i,j} E_{i,j}^{(T)} \leq E_{i}^{(0)},\\
\sum_{j \in V^{(out)}_{i}} q_{i,j} = Q_{i} + \sum_{j \in V^{(in)}_{i} } q_{j,i}, \\
0 \leq q_{i,j}\leq c_{i,j}, Q_{i} \geq 0, E_{i,j}^{(T)}\geq 0, E_{i}^{(R)}\geq 0, E_{i}^{(Q)}\geq 0, E_{i}^{(0)} \geq 0, \\
\end{array} \right.
\end{equation}
where
$E_{i}^{(0)}$ is the initial energy of the $i$-th node,
$E_{i,j}^{(T)}$ is the energy necessary to send one unit of data between $i$-th and $j$-th nodes,
$E_{i}^{(R)}$ is the energy necessary to receive and
$E_{i}^{(Q)}$ is the energy necessary to generate and process a one unit of data by the $i$-th node.
The amount of data generated by the $i$-th node we denote by $Q_i$.
The data flow matrix $q_{i,j}$ defines the amount of data which is sent between $i$-th and $j$-th nodes
and $c_{i,j}$ is the capacity of the corresponding transmission channel $(i,j)$.
By $V^{(out)}_{i}$ and $V^{(in)}_{i}$, $i\in [1,N]$ we denote neighborhoods of each node.
$V^{(out)}_{i}$ is a set of $L_N$ network nodes to which the $i$-th node can send any data, and
$V^{(in)}_{i}$ is a set of nodes from which it can receive any data.
The third equation in (\ref{minimaxNP}) is data flow conservation constraint, which states that the amount of data generated by
each node and the amount of data received from other nodes must be equal to the amount of data which the node can send.
All the matrices $q$, which satisfy the third equation in (\ref{minimaxNP}) form a domain of the objective function of the
problem
\begin{equation} \label{ObjectiveFunction}
E^{*}(q)=\max_{i} \{ E_{i}^{s} \}_{i=1}^{N}.
\end{equation}
To solve the problem (\ref{minimaxNP}) it means to find a global minimum of (\ref{ObjectiveFunction}).
The following simple lemma tells that, any minimum of the objective function (\ref{ObjectiveFunction})
is a solution of the problem (\ref{minimaxNP}).\\\\
{\bf  Lemma 1}. Any minimum of the objective function (\ref{ObjectiveFunction}) is a solution of (\ref{minimaxNP}).\\
{\it Proof}.
The feasible set
$$U_F(q)=\{q \in R^{(N+1)^2}_{+} | \sum_{j \in V^{(out)}_{i}} q_{i,j}- \sum_{j \in V^{(in)}_{i} } q_{j,i}- Q_{i} =0, \; i\in [1,N] \},$$
where $R_{+}$ is a set of non-negative real numbers,
is a domain of the objective function $E^{*}(q)$ and it is a linear in $R^{(N+1)^2}_{+}$.
From the linearity of $U_F(q)$ it follows, that any two points $q_0, q_1\in U_F(q)$ can be connected by a line segment and
$\forall_{t\in [0,1]}\; tq_0 +(1-t)q_1 \in U_F(q).$
The objective function (\ref{ObjectiveFunction}) of the maximum lifetime problem (\ref{minimaxNP})
is a piecewise linear in the $q$ variable,
which means that
$E^{*}(tq_0 +(1-t)q_1)= tE^{*}(q_0) +(1-t)E^{*}(q_1).$
Let us assume, that $q_0$ is a minimum of $E^{*}$ and there exits another local minimum $q_1$
such that
$E^{*}(q_0)<E^{*}(q_1),$
then from the linearity of $E^{*}(q)$ it follows that
$\forall_{t\in [0,1]}\;\; E^{*}(tq_0 +(1-t)q_1)= tE^{*}(q_0) +(1-t)E^{*}(q_1) < tE^{*}(q_1) +(1-t)E^{*}(q_1)$
and $\forall_{t\in [0,1]}\;q(t)$ the inequality
$E^{*}(q(t))< E^{*}(q_1)$
holds, which  means that $q_1$ cannot be a minimum of $E^{*}(q)$ in $U_F(q)$
and $q_0$ is a global minimum. $\diamond$

Generally, we are looking for solution of the problem (\ref{minimaxNP}) in real numbers $q_{i,j}\in R_+^{0}$.
But, if we assume that
\begin{equation} \label{lipp}
Q_{i} \in Z_{+}^{}, \;\;\;\;q_{i,j} \in Z_{+}^{0},
 \end{equation}
then we get the mixed integer linear programming problem for a network lifetime.
Let us assume, that $V^{(out)}_{i} \neq \emptyset$,
and the parameters $c_{i,j}$, $E_{i}^{(0)}$ are sufficiently large,
that at least one solution of (\ref{minimaxNP}) exits. Then we have the following\\\\
{\bf Proposition 1}. The mixed integer linear programming problem for the maximum network lifetime is NP-hard.\\
{\it Proof.}
We reduce the 'partition' problem, \cite{Garey}, to the maximum network lifetime problem given by the set of relations
(\ref{minimaxNP}) and requirement (\ref{lipp}).
Let us write the definition of the feasible set, given by the third relation in (\ref{minimaxNP}),
in the form
\begin{equation} \label{FeasibleSet}
\forall_{i\in [1,N]}\;\;\;\sum_{q_{i,j} \in A_{i}^{'}} q_{i,j} = Q_{i} + \sum_{q_{j,i} \in A_{i} \setminus A_{i}^{'} } q_{j,i},
\end{equation}
where $A_{i} =\{ q_{i,0}, q_{i,j}, q^{}_{j,i} \}_{j=1}^{N}$ and
$A_{i}^{'} = \{ q_{i,0}, q_{i,j} \}_{j=1}^{N}$.
Because the solution $q_{i,j}$ of (\ref{minimaxNP}) must lie on the hyperplane (\ref{FeasibleSet}),
then solving this problem requires to solve the partition problem. $\diamond$

As can be seen from (\ref{minimaxNP}), the problem depends on the topology of the network
$\{ V^{(out)}_{i} \}_{i\in [1,N]}$ and the set of the following parameters
$$ \{ c_{i,j}, Q_{i}, E_{i}^{(R)}, E_{i}^{(Q)}, E_{i,j}^{(T)}, E_{i}^{(0)}\}.$$
Because such optimization problem is to complicated we simplify it,
and consider (\ref{minimaxNP}) only with the following set of parameters
$$ \{ Q_{i}, E_{i,j}^{(T)} \}, \;\; i\in [1,N], j\in [0,N].$$
Which means, that the energy necessary to receive $E_{i}^{(R)}$ and the energy $E_{i}^{(Q)}$ necessary
to generate the amount of $Q_i$  data are equal to zero.
We assume, that the initial energy of each node $E_{i}^{(0)}$ and
the capacity of transmission channels $c_{i,j}$ are sufficiently large that allow the nodes
to send all generated data $Q_i$ to the data collector. We also assume,
that $\forall i\in [1,N], V^{(out)}_{i} = L_N$ which means that any node of the network can send its data to any other node of $L_N$ or to the data collector.
The reduced maximum network lifetime problem, which we are going to investigate, can be given by set of relations
\begin{equation} \label{MinimaxReduced}\left\{   \begin{array}{l}
\min_{q} \max_{i\in L_N} \{ E_{i}^{} \}, \\
E_{i}^{} = \sum_{j=0, j\neq i}^{N} q_{i,j} E_{i,j}^{},\\
h_i(q)-Q_{i}=0,\\
q_{i,j} \geq 0, \;\; Q_{i} > 0, E_{i,j} \geq 0,
\end{array} \right.
\end{equation}
where
$h_{i}(q) = q_{i,0} + \sum_{j=1,j\neq i}^{N} (q_{i,j} - q_{j,i})$, $i\in [1,N]$.

In the next sections we investigate the criteria for which an equal energy solution of the problem (\ref{MinimaxReduced})
exits and give an analytical solution of the problem.
We also define the stability region $U(Q^0, d^0)$ for the $Q$ and $d$ parameters
for which the equal energy solution of (\ref{MinimaxReduced}) can be continuously extended.
\section{Solution of the maximum network lifetime problem for $L_{N}$ }
For any data transmission cost energy matrix of the form (\ref{Eij-a-n-lambda-n}) and wide range
of values of the $Q_i$ parameters, $i\in [1,N]$
there exists an equal energy solution $q^{}$ of the problem (\ref{MinimaxReduced}), i.e.,
a solution for which the energies of each
node are equal $\forall i\in [1,N]\; E_{i}^{s}=E_{N+1-i}^{s}$.
The following lemma estimates the scope of $E_{i}^{s}$ for such solutions.\\\\
{\bf  Lemma 2}.
For the maximum network lifetime problem (\ref{MinimaxReduced}) with $E_{i,j}$ of the form (\ref{Eij-a-n-lambda-n}),
for which there exists an equal energy solution $q^{}(Q)$,
the energy of each node $E_{i}^{s}(q^{})$ is in the interval
\begin{equation} \label{EiScopeAnyaLN}
E_{i}^{s}(q^{}) \in [\frac{1}{N} \sum_{j=1}^{N} j Q_j, \;\sum_{j=1}^{N}Q_j).
\end{equation}
{\it Proof.}
The upper bound of the node energy $E_{i}^{s}(q)$ we estimate for $E_{i,j}=|i-j|^a$, $a\geq 1$.
From the linearity of the problem (\ref{MinimaxReduced}) it follows that this estimation is valid for any $E_{i,j}$
of the form (\ref{Eij-a-n-lambda-n}).
Because for any $a\geq 1$ the energies of the nodes must be finite $E_{i}^{s}(q,a) <\infty$, then we have
$$\forall_{i \in [1,N]}\;\lim_{a \rightarrow \infty} E_{i}^{s}(q,a)= \lim_{a \rightarrow \infty} \sum_{j=0}^{N} q_{i,j} |i-j|^a = q_{i,i-1} + q_{i,i+1}.$$
We show that, for $a>>1$ and any $Q_i\geq 0$, $q_{i,i+1}\rightarrow 0$.
Let us assume that $q_{i,i+1}\neq 0$, because of the 'no loop' requirement
$$q_{i,j}\neq 0 \Rightarrow q_{j,i}=0,$$
the minimum energy of the $(i+1)$-th node necessary to send the amount $(Q_{i+1}+q_{i,i+1})$ of data
in the data collector direction is achieved when the data is transmitted to the $(i-1)$-th node,
omitting the $i$-th node. It means that for $a>>1$ this energy tends to infinity
$$\lim_{a\rightarrow \infty} (Q_{i+1}+q_{i,i+1}+...)E_{i+1, i-1} = (Q_{i+1}+q_{i,i+1}+...)2^a= \infty.$$
Because, in the limit $a\rightarrow \infty$, for any $Q_i\geq 0$,
the solution of (\ref{MinimaxReduced}) tends to the
'next hop' data transmission strategy, then $\max E_{i}^{s} < \sum_{j=1}^{N}Q_{j}$,
which is the upper bound from (\ref{EiScopeAnyaLN}).
To estimate the lower bound of $E_{i}^{s}$ we remark,
that the minimum energy $E_{p}^{\min}(Q_i)$ necessary to deliver the amount $Q_i$ of data to the data collector, for
$E_{i,j}$ of the form (\ref{Eij-a-n-lambda-n}), is achieved for the 'next hop' data transmission strategy $q^{{\rm nh}}$
along the shortest path between the $i$-th node and the data collector, i.e.,
\begin{equation} \label{EpMinNxHop}
E_{p}^{\min}(Q_i,q^{{\rm nh}})= Q_i \sum_{j=1}^{i} E_{j,j-1} = iQ_i.
\end{equation}
The total energy $\sum_{i=1}^{N} E^{s}_{i}(q)$ consumed by all nodes of the $L_N$ network,
for any data transmission flow matrix $q$ is equal to the cost of delivery of all $Q_i$ to the data collector
$$\sum_{i=1}^{N} E_{i}^{s}(q) = \sum_{i=1}^{N} E_{p}^{}(Q_i,q).$$
For an equal energy solutions $q^{}$, $\forall_{i\in [1,N]}\;E^{s}_{i}(q)=E^{s}_{0}$,
we have
$E^{s}_{0} = \frac{1}{N}\sum_{i=1}^{N} E_{p}^{}(Q_i,q)$,
and from (\ref{EpMinNxHop}) it follows that the lower bound of $E_{i}^{s}$ is $\frac{1}{N}\sum_{j=1}^{i} jQ_j$. $\diamond$

The following proposition estimates values of the parameters $Q_i$
for which there exists an equal energy solution of (\ref{MinimaxReduced}) for the
regular ad-hoc wireless network $L_N$.\\\\
{\bf  Proposition 2}.
For the maximum network lifetime problem (\ref{MinimaxReduced}) with $E_{i,j}$ satisfying (\ref{xi-xk-xj}) and $Q_i$
such that
\begin{equation} \label{ConstrainsOnQ}
\forall_{i\in [1,N-1]}\;\;\;Q_{i+1} > \frac{1}{E_{i+1,0}} E_{i}^{s}(Q,q,L_i),
\end{equation}
$Q_1\geq 1$, there  exists an equal energy solution $q$,
and the only non-zero elements of the data flow matrix $q$ are $q_{i,0}$ and $q_{i,i-1}$, $i\in [1,N]$.\\
{\it Proof.}
Let us assume that for $L_N$ network and given set of parameters $Q=(Q_1,...,,Q_N)$ there exits an equal energy solution $q$ of (\ref{MinimaxReduced}),
i.e., a solution for which $\forall_{i\in [1,N]}\;E_{i}^{s}(q)=E_{N+1-i}^{s}(q)$.
We denote the equal energy spend by each node $E^{s}_{i}(Q,L_N)$ by $E^{s}_0(Q,L_N)$.
We add at the end of the $L_N$ network the $(N+1)$-th node and assume that
it generates the amount $Q_{N+1} >\frac{1}{E_{N,0}}$ of data.
If the $(N+1)$-th node sends the amount
$q_{N+1,0}=\frac{1}{E_{N+1,0}}E^{s}_0(Q,L_N)$
of data directly to the data collector,
and the rest of the data
$q_{N+1,N} = Q_{N+1} - \frac{1}{E_{N+1,0}}E^{s}_0(Q,L_N)$
by the 'next hop' data transmission strategy to the data collector,
then it spends the amount
\begin{equation} \label{NodeEnergyN+1}
E^{s}_{N+1}(Q)= Q_{N+1} + (1 - \frac{1}{E_{N+1,0}}) E^{s}_0(Q,L_N)
 \end{equation}
of energy.
Because of the requirement (\ref{ConstrainsOnQ}), the inequality $q_{N+1,N}>0$ is satisfied
and it blocks the data transmission to the $(N+1)$-th node from the nodes of the $L_N$ network.
This means that the energy spend by $(N+1)$-th node $E^{s}_{N+1}(Q,L_{N+1})\geq E^{s}_0(Q,L_{N})$.
The energy $E^{s}_{N+1}(Q,L_{N+1})$ given by the formula (\ref{NodeEnergyN+1}) is optimal because
it is a sum of the minimal energy $E^{s}_0(Q,L_{N})$ which must spend the $(N+1)$-th node
and the energy of the most efficient data transmission in $L_{N+1}$ network - the 'next hop' data transmission strategy. $\diamond$

From the Proposition 2 it follows that for an equal energy solution of (\ref{MinimaxReduced}) in the $L_N$ network,
the energy spend by each node $E^{s}_0(Q,L_{N})$ satisfies the recurrence equation
\begin{equation} \label{NodeEnergyRecursions}
E^{s}_{0}(Q, L_N)= Q_{N} + (1 - \frac{1}{E_{N,0}}) E^{s}_0(Q,L_{N-1}), \;\;N\geq 2,
\end{equation}
where $E^{s}_0(Q,L_{1})= Q_1$.
The equation (\ref{NodeEnergyRecursions}) can be easily solved.
\\\\
{\bf  Lemma 3}. Solution of the recurrence equation (\ref{NodeEnergyRecursions})
for the energy of a node $E^{s}_{0}(L_{N})$
is given by the following polynomial
\begin{equation} \label{NodeEnergyRecursionsSol}
E^{s}_{0}(Q, L_N)  =  Q_{N}  + \sum_{j=2}^{N} Q_{j-1} \prod_{r=j}^{N}(1-\frac{1}{E_{r}}), \;\;N\geq 2,
\end{equation}
where $E_{r}$, $r\in [2,N]$ is the data transmission cost energy matrix $E_{i,j}(\bar{a},\bar{\lambda})$
from (\ref{Eij-a-n-lambda-n}) for $x_i=i$ and $r=i-j$.\\
{\it Proof.} By inserting the formula (\ref{NodeEnergyRecursionsSol}) into (\ref{NodeEnergyRecursions}) we get the
thesis of the lemma. $\diamond$


From the Proposition 2 it also follows that for the amount of data $Q_i$ generated by each node
in $L_N$ network satisfying (\ref{ConstrainsOnQ}),
the only non-zero elements of the optimal data transmission flow matrix $q$ are $q_{i,0}$ and $q_{i,i-1}$.
Putting this requirements into (\ref{MinimaxReduced}) we get the following set of linear, recurrence equations
\begin{equation} \label{qijRecurrenceEq} \left\{   \begin{array}{l}
q_{1,0} = q_{2,0} E_{2} + q_{2,1},\\
q_{i,0} E_{i} + q_{i,i-1} = q_{i+1,0} E_{i+1} + q_{i+1,i}\;\;i \in [2,N],\\
q_{1,0}  = Q_{1} + q_{2,1},\\
q_{i,0} + q_{i,i-1} = Q_{i} + q_{i+1,i}, \;\;i \in [2,N],
 \end{array} \right.\end{equation}
where $E_{r}$ is the data transmission cost energy matrix from the Lemma 3.
The first two equations are the equal energy solution requirements and the next two define the feasible set of the problem
(\ref{MinimaxReduced}).
In the next lemma we give the solution of (\ref{qijRecurrenceEq}).\\\\
{\bf  Lemma 4}. Solution of the set of recurrence equations (\ref{qijRecurrenceEq}),
for a constant $Q_{i}$, is given by the following polynomials
\begin{equation} \label{qijRecurrenceSol}\left\{   \begin{array}{l} 
q_{1,0}  =  Q_{N} + \sum_{j=2}^{N} Q_{j-1} \prod_{r=j}^{N}(1-\frac{E_{1}}{E_{r}}), \\
q_{2,0} =\frac{E_{1}}{E_{2}} Q_{1}, \\
q_{i,0} =\frac{E_{1}}{E_{i}} (Q_{i-1}
+ \sum_{j=1}^{i-2} Q_{i-j-1} \prod_{r=1}^{j}(1-\frac{E_{1}}{E_{i-r}})),
 \;\;\;\; i \in [3,N],\\
q_{2,1} = Q_{2} - \frac{E_{1}}{E_{2}} Q_{1}, \;\;\;\; N=2,\\
q_{2,1} = \sum_{k=2}^{N} Q_{k} - \sum_{k=2}^{N} \frac{E_{1}}{E_{k}} Q_{k-1}
- \sum_{k=3}^{N} \frac{E_{1}}{E_{k}} \sum_{j=1}^{k-2} Q_{k-j-1} \prod_{r=1}^{j} (1-\frac{E_{1}}{E_{k-r}}),
\;\;\;\; N\geq 3,\\
q_{i,i-1} = \sum_{k=i}^{N} Q_{k} - \sum_{k=i}^{N} \frac{E_{1}}{E_{k}} ( Q_{k-1}
 + \sum_{j=1}^{k-2} Q_{k-j-1} \prod_{r=1}^{j} (1-\frac{E_{1}}{E_{k-r}})),
\;\;\;\; i \in [3,N],
\end{array} \right.
\end{equation}
where $E_{r}$ is given by (\ref{Eij-a-n-lambda-n}) for $x_i=i$, and $i-j=r$.\\
{\it Proof.} By inserting the formulas (\ref{qijRecurrenceSol}) into (\ref{qijRecurrenceEq}) we get the
thesis of the lemma. $\diamond$

A special case of (\ref{qijRecurrenceSol}) is a solution the maximum network lifetime problem discussed in \cite{Giridhar}.
For $E_{i,j} = \sum_{s=0}^{\infty}\lambda_{s} |i-j|^{a_s}$ and
$\lambda_s= E_t \gamma^s \frac{1}{s!}$, $a_s = \alpha + s$, $a\geq 2$, $ \gamma \geq 0$,
we obtain the cost energy matrix
$E_{i,j} = E_t |i-j|^{\alpha} e^{\gamma |i-j|}$ from \cite{Giridhar}.
For $E_{i,j}=|i-j|^a$ and $a=2$ the formula for $q_{i,0}$ in (\ref{qijRecurrenceSol}) can be reduced
to the following simple form
$q_{i,0} = \frac{i-H_{i}}{i(i-1)}$,
where $H_i = \sum_{k=1}^{i}\frac{1}{k}$ is the $i$-th harmonic number, \cite{Cichon}.

The next two lemmas describe the size of the stability region $U(Q)$,
for any $Q$ satisfying (\ref{ConstrainsOnQ}), in which the functions (\ref{qijRecurrenceSol}) are solution of  (\ref{MinimaxReduced}).
From the Lemma 3 and Lemma 4 it follows\\\\
{\bf  Lemma 5}.
The solution of the maximum lifetime problem (\ref{MinimaxReduced}) given by (\ref{qijRecurrenceSol})
can be continuously extended to any value of
\begin{equation} \label{QNScope}
Q_N\in (Q_N^{\min}, \infty),
\end{equation}
where
\begin{equation} \label{QNmin}
Q_{N}^{\min} = \frac{1}{E_{N}} (Q_{N-1} + \sum_{j=2}^{N-1} Q_{j-1} \prod_{r=j}^{N-1}(1-\frac{1}{E_{r}}))
\end{equation}
is a solution of the equation $q_{N,N-1}(Q_N^{\min}, L_N)=0$.\\
{\it Proof.} The upper bound for $Q_N$ in (\ref{QNScope}) follows from the Lemma 3 and (\ref{ConstrainsOnQ}).
The lower bound of $Q_N^{}$ is determined by the energy $E_{i}^{s}(L_{N-1})$
consumed by each node in the $L_{N-1}$ network.
If the $Q_N^{}$ value is such that $Q_N^{\min}E_{N,0}=E_{i}^{s}(L_{N-1})$ then
the $N$-th node can send all of its data directly to the
data collector. This means that for $Q_N> Q_N^{\min}$, $q_{N,N-1}(Q_N, L_N)>0$.
This inequality is equivalent to (\ref{ConstrainsOnQ}) for $i=N-1$.
The solution of the equation $q_{N,N-1}(Q_N^{\min}, L_N)=0$, for fixed values of $Q_i$, $i\in [1,N-1]$ is given by (\ref{QNmin}). $\diamond$

The next lemma describes the scope of the parameters $Q_i$, $i\in [1,N-1]$ for which the functions (\ref{qijRecurrenceSol})
are solutions of the maximum network lifetime problem  (\ref{MinimaxReduced}).\\\\
{\bf Lemma 6}. The solution of the maximum network lifetime problem (\ref{MinimaxReduced}) given by (\ref{qijRecurrenceSol})
can be continuously extended to any value of
\begin{equation} \label{QiScope}
 Q_i\in (0, Q_i^{\max}), \;\;i\in [1,N-1],
\end{equation}
where $Q_i^{\max}$ is a solution of the equation
\begin{equation} \label{qi+1iEq}
q_{i+1,i}(Q_i^{\max}, L_N)=0, \;\;i\in [1,N-1].
\end{equation}
{\it Proof.}
The constrains (\ref{ConstrainsOnQ}) give the lower bound of $Q_i$ for which
a solution of (\ref{MinimaxReduced}) is given by (\ref{qijRecurrenceSol}).
The constrains were obtained by calculating recursively the minimum value of $Q_i$ for the $L_i$ network,
for which the $i$-th node is the last one and does not receive any data from other nodes.
We can improve the estimation of the lower bound for $Q_i$ by taking into account the data received by the $i$-th node
from nodes $j\in [i+1, N]$ of the $L_N$ network.
The improved estimation of $Q_i$, $i\in [1,N-1]$ for the $L_N$ network is given by the following set of
inequalities
\begin{equation} \label{ConstrainsOnQWeak}
\left\{   \begin{array}{l}
Q_{1} + q_{2,1}(Q_{1},L_N) > 0, \\
(Q_{i} + q_{i+1,i}(Q_i,L_N)) > \frac{1}{E_{i}} E_{0}^{s}(Q,q,L_{i-1}),\;\;i\in [2,N-1],
\end{array} \right.
\end{equation}
where $q_{i+1,i}(Q,L_N)$ is given in (\ref{qijRecurrenceSol}).
The above inequalities allow us to determine the maximum value of $Q_{i}$ for which the functions
(\ref{qijRecurrenceSol}) are solution of (\ref{MinimaxReduced}).
Instead of minimizing the $Q_{i+1}$ value in (\ref{ConstrainsOnQWeak})
we can maximize $Q_{i}$ and thus the energy $E_{0}^{s}(L_{i})$, such that $q_{i+1,i}=0$.
The requirement for the maximum value of $Q_{i}$, $i\in [1,N-1]$ is given by the set of equations
\begin{equation} \label{QiMaxInequality} \left\{   \begin{array}{l}
(Q_2 + q_{3,2}(L_N))\; E_{2} = Q_{1}^{\max},\\
(Q_{i+1} + q_{i+2,i+1}(Q_i^{\max},L_N))\;E_{i+1}  = E_{0}^{s}(Q_i^{\max},L_{i}),\;\;i\in [2,N-2],\\
Q_N \; E_{N} = E_{0}^{s}(Q_{N-1}^{\max}, L_{N-1}),
\end{array} \right.\end{equation}
where $E_{0}^{s}(Q_i^{\max},L_{i})=q_{1,0}(Q_i^{\max}, L_{i})$.
With the help of (\ref{qijRecurrenceSol}) one can easily check that the formulas
(\ref{QiMaxInequality}) and (\ref{qi+1iEq}) are equivalent.
To estimate the lower bound of $Q_{i}$, $i\in [2,N-1]$ we use the inequalities (\ref{ConstrainsOnQWeak}).
The inequalities (\ref{ConstrainsOnQWeak}) are equivalent to set of the equations for $Q_i^{\min}$
$$q_{i,i-1}(Q_i^{\min}, L_N)=0,\;\; i\in [2,N-1],$$
which means that all of data of the $i$-th node is sent directly to the data collector.
From the first inequality in (\ref{QiMaxInequality}) it follows that $\forall_{Q_1}\;q_{2,1}(Q)>0$
and $Q_1$ can be arbitrary small, which means that $Q_1^{\min}=0$.
%
%
From (\ref{qijRecurrenceSol}) it follows that the inequality $q_{i,i-1}(Q_{i}, L_N)>0$
is satisfied for arbitrary small $Q_i$, and thus $Q_i^{\min}=0$. $\diamond$
\section{Solution stability under modification of node location}
Let us denote by $S_N$ a one-dimensional, ad-hoc wireless network
which nodes are located at the points $x_i=i-d_i$ of the half-line, where
$d_i \in (-1,1)$, $i \in [1,N]$.
We know, that the equal energy solution of (\ref{MinimaxReduced}) with the data transmission cost energy matrix (\ref{Eij-a-n-lambda-n})
for $d_i=0$ is given by (\ref{qijRecurrenceSol}).
We expect, that for small $d_i$ the solution of (\ref{MinimaxReduced}) will be also of the form (\ref{qijRecurrenceSol}).
The solution (\ref{qijRecurrenceSol}) has been obtained under assumption that $\forall_{i\in [1,N]} E_{i,i-1}=1$.
Change of the node location causes change of the data transmission cost energy matrix elements $E_{i,i-1}\neq 1$
and the set of functions (\ref{qijRecurrenceSol}) must be modified to get proper solution
of (\ref{qijRecurrenceEq}) with $E_{i,i-1}\neq 1$.
The set of linear equations (\ref{qijRecurrenceEq}) for an equal energy solution of (\ref{MinimaxReduced})
in the $S_N$ network we write in the matrix form.
The data conservation flow $h_{}(q)=Q_0$ constraints and the equal energy solution requirements $E_{i+1}=E_{i}$, $i \in [1,N-1],$
we write in the form
\begin{equation} \label{Mq=Q}
M^{} \vec{q} = \vec{Q}_{0}, \end{equation}
where
$\vec{q} = (q_{N,0}, ..., q_{1,0}, q_{2,1}, ...,q_{N,N-1})$,
$\vec{Q}_{0} = (1, ...,1, 0, ... ,0)$,
and
$$M=\left(
   \begin{array}{cc}
     I_{N,N} & A_{N,N-1} \\
     B_{N-1,N} & C_{N-1,N-1}
   \end{array}
 \right)$$
is an invertible matrix $(2N-1)\times (2N-1)$.
The elements of the matrix M have the form
\begin{equation} \label{IABC}
\left\{   \begin{array}{ll}
I_{i,j} = \delta_{i,j},                                         & i,j \in [1,N], \\
A_{i,j} = \delta_{i,N-j} - \delta_{i,N-j+1},                    & i \in [1,N], j \in [1,N-1],\\
B_{i,j} = - E_{i+1,0}\; \delta_{i,N-j} +   E_{i,0}\; \delta_{i,N-j+1},  & i \in [1,N-1], j \in [1,N],\\
C_{i,j} = - E_{i+1,i}\; \delta_{i,j} + E_{i,i-1}\; \delta_{i,j+1},      & i,j \in [1,N-1].
\end{array} \right.
\end{equation}
In (\ref{IABC}) the data transmission cost energy matrix $E_{i,j}$ is given by (\ref{Eij-a-n-lambda-n}).
The form of the matrix $M$ is chosen in such way that allows in an easy way to extend it
for an arbitrary number of nodes added to the $S_N$ network.
Namely, after adding a new node to $S_N$ network we must resize the matrix $M$
by adding two rows, one on the top and one on the bottom of $M$,
and two columns, one on the left hand side, and one on the right hand side of $M$.
In this section we would like to investigate the stability of the problem (\ref{MinimaxReduced})
under modification of node location by change of the parameters $d=(d_1, ...,d_N)$.
We estimate the size of the intervals $d_i\in (d_i^{L},d_i^{R})$
for which the solution (\ref{Mq=Q}) of (\ref{MinimaxReduced}) can be extended, where
$d_i^{L}\in (-1,0)$ and $d_i^{R}\in (0,1)$ are maximum shifts from the initial location $x_i=i$ of the node
to the left and right respectively.
In other words, we will search for a networks $S_N$ with the data transmission cost energy matrix (\ref{Eij-a-n-lambda-n})
for which the solution of the problem (\ref{MinimaxReduced}) satisfies (\ref{Mq=Q}).
To estimate the asymptotic properties of the optimal energy spend by each node of the $S_N$ network
for the maximum network lifetime problem we must fix the size of the network $D(S_N)=|x_N-x_0|$.
The following lemma generalizes the Lemma 2 and estimates the scope of
the energy spend by each node of the $S_N$ network, when $E_{i,j}$ is of the form (\ref{Eij-a-n-lambda-n}).\\\\
{\bf  Lemma 7}. For the maximum network lifetime problem (\ref{MinimaxReduced}), and $E_{i,j}$ of the form (\ref{Eij-a-n-lambda-n}),
for which there exists an equal energy solution $q^{}(Q)$,
the energy of each node $E_{i}^{s}(q^{})$ is in the interval
\begin{equation} \label{EiScopeAnyaSN}
E_{i}^{s}(q^{}) \in [\frac{1}{N}\sum_{i=1}^{N} Q_i \sum_{j=1}^{i} E_{j,j-1},
\; \max_{i\in[1,N]} \{ E_{i,i-1} \sum_{j=i}^{N}Q_j \}).
\end{equation}
{\it Proof.} The lower bound for $E_{i}^{s}(q^{},Q)$ follows from the property (\ref{xi-xk-xj})
of the data transmission cost energy matrix (\ref{Eij-a-n-lambda-n}) and the Lemma 2.
The upper bound we determine for $E_{i,j}=|x_i-x_j|^a$ where $a\geq 1$. From linearity of the problem
it follows that it is valid for any data transmission cost energy matrix of the form (\ref{Eij-a-n-lambda-n}).
We can use the result of the Lemma 2, which states that for $a \rightarrow \infty$ the optimal transmission
tends the 'next hop' data transmission strategy.
For a fixed size of the network $D(S_N)=|x_N-x_0|=N$ and nodes location at points $x_i=i-d_i$, $d_i\in(-1,1)$,
there is at least one distance $d_{i,j}=|x_i-x_j|>1$. For the $i$-th node which distance
to the neighboring node is greater than 1, the cost of data transmission does not exceed $E_{i,i-1} \sum_{j=i}^{N}Q_j$ and
for $a \rightarrow \infty$ grows to infinity. The maximum node cost $E_{i}^{s}$ must be selected among all values
$E_{i,i-1} \sum_{j=i}^{N}Q_j$ for  $i\in [1,N]$, what is stated in the formula (\ref{EiScopeAnyaSN}). $\diamond$

The next lemma gives the exact form of the energy of each node $E_{i}^s(q,S_N)$, $i\in [1,N]$ for an equal energy solution of
(\ref{MinimaxReduced})  for $S_N$ network and it is generalization of the formula (\ref{NodeEnergyRecursionsSol}).\\\\
{\bf  Lemma 8}. The energy $E_{0}^s(S_N)$  of each node of the $S_N$ for an equal energy solutions
of (\ref{Mq=Q}) is given by
\begin{equation} \label{SolutionOfRecursionFor-E0-SN}
E_{0}^s(S_N)= \frac{1}{\det M_N} \sum_{k=1}^{N-1} Q_k \prod_{i=1}^{k} E_{i, 0} \; \prod_{i=k+1}^{N}(E_{i, 0} - E_{i,i-1}) + Q_N \prod_{i=1}^{N} E_{i, 0}, \;\;N\geq 2,
\end{equation}
where $E_{0}^s(L_{1})= E_{1,0}$.\\
{\it Proof.}
For an equal energy solution $E_{i}^s(S_N)=E_{N-i+1}^s(S_N)$, $i\in [1,N]$,
the equation (\ref{Mq=Q}) for $q_{1,0}$
has the form
\begin{equation} \label{RecursionFor-E0-SN}
E_{0}^s(S_N) = Q_N \frac{\prod_{i=1}^{N}E_{i, 0}}{\det M_N} + (E_{N, 0} - E_{N,N-1}) \frac{\det M_{N-1}}{\det M_N}E_{1}^s(L_{N-1}),
\end{equation}
where $E_{0}^s(S_N) = E_{i}^s(S_N)$, $i\in [1,N]$.
It is easy to check that $E_{0}^s(S_N)$ given by (\ref{SolutionOfRecursionFor-E0-SN})
is a solution of (\ref{RecursionFor-E0-SN}) with the initial condition $E_{0}^s(L_{1})= E_{1,0}$. $\diamond$

The general solution of the equations (\ref{Mq=Q}) has very complicated form.
In this paper, we investigate the stability of the solution of the equation (\ref{Mq=Q})
under modification of each parameter $d_{i}\in (-1,1)$ separately and for $Q_i=1$, $i\in [1,N]$.
It means that, we will change the $i$-th node location,
keeping all other nodes of $S_N$ in the points $\forall_{j \neq i}\; x_{j}=j$, i.e.,  $\forall_{j \neq i} \;d_{j}=0$.
The necessary requirement for the stability of a solution (\ref{Mq=Q}) with respect to the parameter
$d_i$ is given by the set of following inequalities $ q_{i,0}(d,S_N)>0, i\in [1,N]$,
where $(d_1, ...,d_N)\in (-1,1)^{N}$.
The maximal possible shift of the $i$-th node to the left $d_{i}^{L}$ and to the right $d_{i}^{L}$
from the initial location $x_i=i$ satisfies the following equations
$$\left\{   \begin{array}{l}
q_{i,0}(d_{i}^{L},S_N)=0,\\
q_{i+1,0}(d_{{i}}^{R},S_N)=0.
\end{array} \right.$$ 
The next proposition estimates the scope of $d_{i}$ for any $E_{i,j}$ of the form (\ref{Eij-a-n-lambda-n}),
for which the node can be moved without changing the optimal transmission strategy.
We have the following\\\\
{\bf Proposition 3}.
For the parameter $d_{i}$, $i\in [1,N]$, $N>3$, $Q_j=1$, $j\in[1,N]$,
the stability region of $U(d_i,E)$, for any $E_{i,j}$ of the form (\ref{Eij-a-n-lambda-n}),
is inside the interval
$$U(d_i,E) \subseteq (d_{i}^{L},  d_{i}^{R}),$$
where
\begin{equation} \label{Scope-dLR}
\left\{   \begin{array}{l}
d_1^L(L_N) = -1, \\
d_i^L(L_N) =  - \frac{1}{4} (\sqrt{N(8 i + N(N+1 - i)^2)} - N(N+1 - i)),\;\; i < \frac{ N^2+N+2}{2N}, \;\; i\in [2,N-1], \\
d_i^L(L_N)=-1,  {\rm otherwise}, \\ 
d_N^L(L_N) = -1, \\\\
d_1^{R}(L_N) = \frac{1}{2 N - 4} (\sqrt{N (N^3 + 2 N^2 + 5 N - 8)} - N - N^2), \;\; N>2,\; d_1^{R}(L_2)=\frac{1}{3}, \\
d_i^{R}(L_N) = \frac{1}{ 4 (N-1 - i)}(\sqrt{ N (8 i (1 + i)(N-1 - i) + N (3 - i^2 + N + i N)^2)} - N^2(1+i)+ N(i^2-3)),\\
i < \frac{1}{4N} ( N^2-N-2 + \sqrt{4 - 12 N + 37 N^2 + 6 N^3 + N^4}), \;\; N > 3, i\in [2,N-2]\\
d_i^{R}(L_N)=1, {\rm otherwise}, \\
d_{N-1}^R(L_N) = 1,  d_N^R(L_N) = 1.
\end{array} \right.\end{equation}
{\it Proof}.
Because the stability region $U(d_i,E)$ for any $E$ of the form (\ref{Eij-a-n-lambda-n}) is smaller than
the stability region $U_0(d_i)$ for $E_{i,j}=|x_i-x_j|$, then it is enough to show that $U_0(d_i)$ is given by (\ref{Scope-dLR}).
The solution of the equation (\ref{Mq=Q}) for $E_{i,j}=|x_i-x_j|$, where $x_i=i-d_i$ has the form
\begin{equation} \label{04}\left\{   \begin{array}{l}
 q_{i,0}(d_{i},S_N)  = \frac{1}{2N(i-d_i)} (i N + (N + 1 - i)N d_i - 2 d_i^2),\\
 q_{i+1,0}(d_{i},S_N) = \frac{1}{2(i+1)N(i-d_i)} (i(i+1)N - ( N(i+1 ) - i^2 + 3)N d_i - 2 (N-1-i) d_i^2).
\end{array} \right.
\end{equation}
The roots of (\ref{04}) with respect to $d_i$ variable are given by (\ref{Scope-dLR}).$\diamond$

The Figure 1 shows the allowed moves of the $i$-th sensor for which the data flow matrix $q(d_i)$
is a solution of the maximum lifetime problem given by (\ref{Mq=Q}).

\begin{figure}[!ht]
\begin{center}
\includegraphics{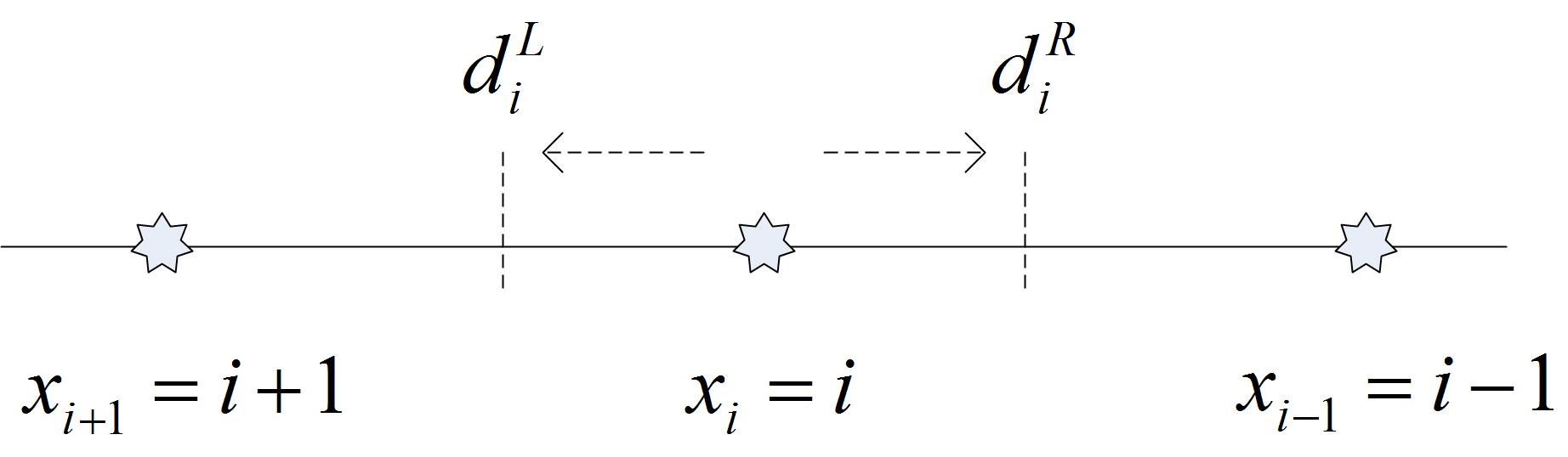}
\caption{Moves of the $i$-th node under which the solution $q(d_i)$ is stable.}
\label{Fig01di}
\end{center}
\end{figure}

From the Proposition 3 it follows that when the number of nodes grows, then the stability regions $U(d_i,E)$ shrinks.
The exact size of the stability regions $U(d_i,E)$ can be calculated by solving the
equations (\ref{Mq=Q}) for particular form of (\ref{Eij-a-n-lambda-n}) and finding the interval in which
inequalities $q_{i,0}(d_{i}^{},S_N)>0,$ and $q_{i+1,0}(d_{{i}}^{},S_N)>0$ are satisfied.
For example, for $a=1$, we have $-1 < d_1(S_3) < 0.245$,
but for $a=1.1$, $-1 < d_1(S_3) < 0.21 $ or for $a=2$,  $-1 < d_1(S_3)  < 0.1 $.
\section{Conclusions}
In the paper we solved the maximum lifetime problem
for a one-dimensional, regular ad-hoc wireless network with one data collector $L_N$
for any data transmission cost energy matrix which elements $E_{i,j}$ are
superadditive functions, i.e., satisfy the inequality $\forall_{i\leq j\leq k}\;E_{i,j}+E_{j,k}\leq E_{i,k}$.
We analyzed stability of the solution under modification of two sets of parameters,
the amount of data $Q_i$, $i\in [1,N]$ generated by each node
and location of the nodes $x_i$ in the network.
Because we assumed, that the data transmission cost energy matrix $E_{i,j}$ is a function of a distance between network nodes
then change of the node location caused the change of $E_{i,j}$.
In the paper we estimated the size of the stability region $U(Q^0,d^0)$
for the solution of the maximum network lifetime problem in $L_N$ network,
where $Q^0_i=1$, $d^0_i=0$
and $d_i\in (-1,1)$ describes the shift of the nodes from their initial location $x_i^0=i$, i.e., $x_i=i-d_i$.
\section*{Appendix 1}
In this appendix we prove by induction that the set of functions (\ref{qijRecurrenceSol}) for  $Q^{0}_{i}=1$
and $E_{i,j}$ of the form (\ref{Eij-a-n-lambda-n}) is a solution of (\ref{MinimaxReduced}).
In the first step, we estimate the size of the stability region $U(Q^{0})$
for which the solution (\ref{Eij-a-n-lambda-n}) of (\ref{MinimaxReduced}) can be continuously extended.\\\\
{\bf  Lemma 1}. The solution $q(Q^0)$ given by (\ref{qijRecurrenceSol}) of (\ref{MinimaxReduced}) for $E_{i,j}=|i-j|$,
where $Q_i^{0}=1$, $i\in [1,N]$
can be continuously extended to any $Q\in U(Q^{0})$, where
\begin{equation} \label{UQSet}
U(Q^{0})=\{Q \in R_{+}^{N}| \sum_{i=1}^{N} Q_i < \frac{3}{2}N, \;Q_i\geq 1, i\in [1,N] \}\;\;N\geq 3.
\end{equation}
{\it Proof.} From the solution of (\ref{qijRecurrenceSol}) for $E_{i,j}=|i-j|$, we can estimate the size of the
interval for $Q_i$, when $Q_{j\neq i}=1$, for which $q_{i+1,i}(Q)>0$.
From (\ref{qijRecurrenceSol}) it follows that
$q_{i+1,i}(Q)>0$ when $Q_{i}\in [1, \frac{1}{2}N + 1)$, $i\in [1,N]$.
These requirements we can write in the form $\sum_{i=1}^{N} Q_{i} < \frac{3}{2}N$, which was used in the definition of the stability region
(\ref{UQSet}). $\diamond$

Let us denote by $U(Q^0,E)$ the stability region defined by all points $Q$ for which the solution (\ref{qijRecurrenceSol})
of (\ref{MinimaxReduced}) can be extended, where $E$ denotes the data transmission cost energy matrix of the form (\ref{Eij-a-n-lambda-n}).
Then we have the following inclusion
\begin{equation} \label{UQInUQE}
U(Q^0) \subseteq U(Q^0,E),
\end{equation}
and both sets are equal for $E_{i,j}=|i-j|$. From the above formula it follows \\\\
{\bf Lemma 2}. For any point $Q$ of the region $U(Q^{0})$ and any data transmission cost energy matrix $E_{i,j}$ of the form (\ref{Eij-a-n-lambda-n})
the data flow matrix given by (\ref{qijRecurrenceSol}) is a solution of (\ref{MinimaxReduced}). $\diamond$
%

With the help of the Lemma 2 we can prove by induction that\\\\
{\bf  Lemma 3}. The set of functions (\ref{qijRecurrenceSol}) for $Q_i^{0}=1$ and arbitrary $E_{i,j}$ of the form (\ref{Eij-a-n-lambda-n})
is an equal energy solution of (\ref{MinimaxReduced}).\\
{\it Proof.}
We assume that the functions in (\ref{qijRecurrenceSol}) are solution of (\ref{MinimaxReduced}) for some N and prove
that they are solution for $N+1$.
When we add the $(N+1)$-th node at the end of the network $L_N$, because of the inclusion (\ref{UQInUQE}) we know that
the transmission of $Q_{N+1}=1$ of data via the $L_N$ network does not change the optimal data transmission
given by the function (\ref{qijRecurrenceSol}).
The equal energy solution for $L_{N+1}$ network means that
\begin{equation} \label{EqualEnergyN_1}
E_{N+1}^{s} = E_{0}^{s}(Q_i^0 + q_{N+1,i}, L_N),
\end{equation}
where
\begin{equation} \label{EnergyN_1}
E_{0}^{s}(Q_i^0+q_{N+1,i},L_N) = E_{0}^{s}(Q^0,L_N) + q_{N+1,N} + \sum_{j=2}^{N} q_{N+1,j-1} \prod_{r=j}^{N}(1-\frac{1}{E_r}),
\end{equation}
$E_{0}^{s}(Q^0,L_N)=q_{1,0}(Q^0)$ is given in (\ref{qijRecurrenceSol}) and $E_r$ is defined in the Lemma 4.
Inserting the equation (\ref{EnergyN_1}) into (\ref{EqualEnergyN_1}), for $Q_i=1$ we get
\begin{equation} \label{ENplus1}
E_{N+1}^{s}= E_{0}^{s}(Q^0,L_N) + q_{N+1,N} + \sum_{j=2}^{N} q_{N+1,j-1} \prod_{r=j}^{N}(1-\frac{1}{E_r}).
 \end{equation}
From the general formula for the $(N+1)$-th node energy
$$E_{N+1}^{s} = q_{N+1,N} + q_{N+1,0}E_{N+1} + \sum_{j=2}^{N} E_{N+2-j} q_{N+1,j-1},$$
and (\ref{ENplus1}) we get the formula for the energy spend by the $(N+1)$-th node to send $q_{N+1,0}$ of data directly to the data collector
\begin{equation} \label{Max-qNN+1}
q_{N+1,0} E_{N+1} = E_{0}^{s}(Q^0,L_N) - \sum_{j=2}^{N} q_{N+1,j-1} (E_{N+2-j} - \prod_{r=j}^{N}(1-\frac{1}{E_r}) ).
 \end{equation}
If this energy is maximal, then the amount of data $Q_{N+1}-q_{N+1,0}$ send via the $L_N$ network is minimal.
Because of the equal energy solution requirement (\ref{EqualEnergyN_1}),
this minimal energy is equal to the $(N+1)$-th node energy $E_{N+1}^{s}$.
If the $(N+1)$-th node will send a non-minimal amount of data via the $L_N$ network
(saving the energy by sending a non-maximal amount of data directly to the data collector),
for example it will send all of the data $Q_{N+1}-q_{N+1,0}$ to some $i$-th node,
then because of the stability of the solution in the $U(Q^0)$ region,
see (\ref{UQInUQE}), this amount of data will increase the energy in the $L_i$ network and the nodes $j\in [N,i+1]$
from the $L_N$ network cannot lower this energy.
From (\ref{Max-qNN+1}) it follows that the maximum value of  $q_{N+1,0}$ we obtain  when
$\forall_{j\in [2,N]}\; q_{N+1,j-1}=0$ and $q_{N+1,0} = \frac{1}{E_{N+1}} E_{0}^{s}(Q^0,L_N)$.
Because $q_{N+1,N} = Q_{N+1}^0 - q_{N+1,0}$, where $Q_{N+1}^0 =1$  we get
the minimum energy of each node
\begin{equation} \label{EN1}
E_{0}^{s}(L_{N+1}) = E_{N+1}^{s} = 1 + (1 - \frac{1}{E_{N+1}}) E_{0}^{s}(L_N),
 \end{equation}
which is equal to $q_{1,0}$ in (\ref{qijRecurrenceSol}) for the $L_{N+1}$ network. $\diamond$
\section*{Appendix 2}
In this appendix we solve the set of linear equations (\ref{Mq=Q}) for
the data transmission cost energy matrix of the form $E_{i,j}=|x_i-x_j|$, where $x_i=i-d_i$.
Solution of (\ref{Mq=Q}) for $d_i$ parameters which satisfy (\ref{Scope-dLR}) is
an equal energy solution of the maximum lifetime problem (\ref{MinimaxReduced}) in the $S_N$ network.
Because for the data transmission cost energy matrix of the form $E_{i,j}=|x_i-x_j|$
the costs of delivery of any data to the data collector do not depend on the transmission strategy along the shortest path,
then there exits an infinite number of solutions of (\ref{MinimaxReduced}).
The particular solution $q$ of (\ref{MinimaxReduced}) for $E_{i,j}=|x_i-x_j|$, which satisfies (\ref{Mq=Q}),
we used in the Proposition 3 to estimate the size of the stability region $U(d_i,E)$
and to estimate the size of $U(Q^0)$, see (\ref{UQInUQE}).
\\\\
{\bf  Lemma 1}. An equal energy solution of the maximum lifetime problem in the $S_N$ network
for the data transmission cost energy matrix $E_{i,j}=|x_i-x_j|$ and $Q_i$ satisfying (\ref{ConstrainsOnQ})
is given by the data flow matrix
\begin{equation} \label{SolutionFor-a-1-any-x-Q}
\left\{   \begin{array}{l}
q_{1,0} = \frac{1}{N}\frac{1}{x_1}\sum_{j=1}^{N} x_j Q_j, \\
q_{i,0}= \frac{1}{N x_{i}x_{i-1}}( N (x_i - x_{i-1}) \sum_{j=1}^{i-1} x_j Q_j  + (i x_{i-1} - (i-1)x_{i}) \sum_{j=1}^{N} x_j Q_j ),\;\; i\in [2,N], \\
q_{N,0} = (1 - \frac{N-1}{N}\frac{x_N}{x_{N-1}}) Q_N  + \frac{1}{N}\frac{1}{x_{N-1}} \sum_{j=1}^{N-1} x_j Q_j, \\
q_{i,i-1} = \frac{1}{N}\frac{1}{x_{i-1}} ((i-1)\sum_{j=i}^{N} x_j Q_j  - (N-i+1)\sum_{j=1}^{i-1} x_j Q_j), \;\;i\in [2,N],
\end{array} \right.\end{equation}
where $x_i=i-d_i$ and $d_i$ satisfies (\ref{Scope-dLR}).\\
{\it Proof.}
If $Q_i$ satisfies (\ref{ConstrainsOnQ}), then the data is transmitted in the $S_N$ network along the shortest path between nodes of the network.
For $E_{i,j}=|x_i-x_j|$ the cost of delivery of $Q_i$ of data to the data collector does not depend on the transmission path,
and it is equal
\begin{equation} \label{PathEnergyQi-a-1}
E_{p}^{\min}(Q_i,q, S_N)= x_i Q_i.
\end{equation}
For an equal energy solution of the maximum network life time problem the energy of each node is equal to
$$E^{s}_{0} = \frac{1}{N}\sum_{j=1}^{N} x_j Q_j.$$
From (\ref{PathEnergyQi-a-1}) and $E^{s}_{0}=E_{1,0}q_{1,0}= x_1 q_{1,0}$ we get the first equation in (\ref{SolutionFor-a-1-any-x-Q}).
From the Proposition 2 it follows that the solution of the maximum network lifetime problem (\ref{MinimaxReduced})
can be reduced to the set of linear equations (\ref{Mq=Q}).
For $E_{i,j}=|x_i-x_j|$ the equations (\ref{Mq=Q}) can be written
in the form
\begin{equation} \label{ConstrainQi-a-1}
\left\{   \begin{array}{l}
x_{1} q_{1,0} = x_{2} q_{2,0} + (x_{2}-x_{1}) q_{2,1},\\
x_{i} q_{i,0} + (x_{i}-x_{i-1}) q_{i,i-1} = x_{i+1} q_{i+1,0} + (x_{i+1}-x_{i}) q_{i+1,i}, \;i\in [2,N-1],\\
q_{1,0} = Q_{1} + q_{2,1},\\
q_{i,0} + q_{i,i-1} = Q_{i} + q_{i+1,i}, \; i\in [2,N-1].
\end{array} \right.\end{equation}
By a direct computation one may check that $q_{i,i-1}$ given by the fourth relation in (\ref{SolutionFor-a-1-any-x-Q})
satisfies the following formula
$$2 x_i q_{i+1,i} - x_{i-1}q_{i,i-1} -  x_{i+1}q_{i+2,i+1} = x_{i+1} Q_{i+1} - x_i Q_i,$$
which follows from (\ref{ConstrainQi-a-1}).
From the relation $q_{i,0} = Q_{i} + q_{i+1,i} - q_{i,i-1}$ we can determine $q_{i,0}$
given in (\ref{SolutionFor-a-1-any-x-Q}).
From the Proposition 3 we know that for $x_i=i-d_i$ and $d_i$ which satisfies (\ref{Scope-dLR})
the set of functions (\ref{SolutionFor-a-1-any-x-Q}) is a solution of the maximum network lifetime problem (\ref{MinimaxReduced}).
$\diamond$
%

\end{document}